\documentclass{emulateapj}
\usepackage{apjfonts}
\usepackage{psfig}

\newcommand\mes{M{\'e}sz{\'a}ros}

\begin{document}

\shorttitle{Universal GRB jets}
\shortauthors{Lazzati \& Begelman}
\title{Universal GRB jets from jet-cocoon interaction in massive
stars}

\author{Davide Lazzati and Mitchell C. Begelman\altaffilmark{1}}
\affil{JILA, 440 UCB, University of Colorado, Boulder, CO 80309-0440,
  USA} 
\email{lazzati;mitchb@colorado.edu} 
\altaffiltext{1}{Department of Astrophysical and Planetary Sciences,
University of Colorado at Boulder}

\begin{abstract}
We consider the time-dependent evolution of a relativistic jet
following its breakout through the surface of a massive compact star,
as envisaged in the collapsar model of gamma-ray bursts.  At breakout,
the jet is tightly collimated by the pressure of its hot cocoon, which
is created as the jet traverses the star.  After breakout, the cocoon
pressure drops and the jet expands toward its natural opening
angle. We show that the evolving opening angle of the jet produces a
stratification of the total energy with the off-axis angle, resulting
in a universal morphology.  The angular structure is largely
independent of the initial beam pattern and depends only on the
luminosity of the central engine.  With the minimal assumption of a
constant energy release we reproduce the $\theta^{-2}$ profile
required to explain observations of afterglows.
\end{abstract}

\section{Introduction}

After the observation of GRB~990123 and its afterglow, it was realized
that Gamma-Ray Bursts (GRBs) are collimated and not isotropic
explosions. GRB~990123 was a most problematic event since it called
for more than a solar mass of energy entirely converted into
$\gamma$-ray photons (Kulkarni et al. 1999) to explain its fluence
under the assumption of isotropic emission. It also had a peculiar
afterglow, with a steepening of the decay rate approximately one day
after explosion (Castro-Tirado et al. 1999). This behavior had been
predicted to be the signature of a collimated outflow (Rhoads 1997,
1999). After correcting for beaming the implied energy release was
reduced to $\sim5\times10^{51}$~erg (Bloom, Frail \& Kulkarni 2003),
well below the energy crisis level of $10^{53}$~erg.

GRB jets were initially postulated to be uniform, i.e., with constant
properties (energy flux, Lorentz factor, pressure) within a
well-defined opening angle $\theta_j$ and nothing outside. Such jets
would produce a steepening afterglow light curve, with the break time
scaling as $\theta_j^{8/3}\,(E_{\rm{iso}}/n)^{1/3}$ (Sari, Piran \&
Halpern 1999), where $E_{\rm{iso}}$ is the isotropic equivalent energy
of the outflow and $n$ the density of the ambient medium. Frail et
al. (2001) and Panaitescu \& Kumar (2001) discovered a remarkable
anti-correlation between the jet break time and the isotropic
equivalent energy release $E_{\gamma,\rm{iso}}$ in the prompt
phase. Such a correlation, initially entirely empirical, can be
accounted for if it is assumed that the same total energy is given to
every GRB but is channeled into jets with different opening angles.  A
narrow jet would appear as a very bright GRB followed by an
early-breaking afterglow, while a wide jet would produce a weak GRB
and an afterglow with a late break.

Rossi, Lazzati \& Rees (2002) realized that the relation
$E_{\gamma,\rm{iso}}\,\theta_j^2=$constant may reflect a universal
angular distribution of jet energy rather than a distribution of
opening angles among different jets (see also Lipunov, Postnov \&
Prokhorov 2001; Zhang \& \mes\ 2002).  In this case one would write
\begin{equation}
\frac{dE(\theta)}{d\Omega} \propto \theta^{-2}.
\label{eq:stru}
\end{equation}
Rossi et al. also showed that such an energy pattern would produce an
afterglow that reproduces the Frail et al. correlation and does not
violate other observations. Different observed properties would
reflect different observing geometries rather than different intrinsic
jet properties. Several other jet morphologies have since been
proposed and studied, specifically general power-law profiles and
Gaussian profiles (Granot \& Kumar 2003; Salmonson 2003, Rossi et
al. 2004; Zhang et al. 2004a). Despite this theoretical interest,
research on structured jets has focused on their late-time dynamics
(Kumar \& Granot 2003) and observable properties (Perna, Sari \& Frail
2003; Nakar, Granot \& Guetta 2004) and not on the origin of the
energy distribution. The only exception is the electromagnetic
force-free model, in which a structured jet is predicted naturally
(Lyutikov, Pariev \& Blandford 2003).

Some degree of structure in GRB jets is implied by their connection to
supernova (SN) explosions (Galama et al. 1998; Stanek et al. 2003;
Hjorth et al. 2003; Malesani et al. 2004). The shear forces and mixing
of the cold stellar material with the outflowing relativistic plasma
are expected to create an interface or hollow slower jet around the
hyper-relativistic flow (MacFadyen \& Woosley 1999; MacFadyen et
al. 2001; Proga et al. 2003). The resulting energy distributions of
the highly relativistic material are, however, far from a power-law
(Zhang, Woosley \& Heger 2004b). If one includes the
trans-relativistic outflow energy a power-law profile is obtained, but
with the steep index $-3$.

In this letter we propose a mechanism to create structured jets with
$dE/d\Omega \propto \theta^{-2}$ as a result of the time evolution of
the opening angle of an hydrodynamic jet. This mechanism relies on the
nature of the jet propagation through the star but is more robust and
predictable than the poorly understood shear effects and mixing
instabilities. We show that, under a simple ansatz (constant
luminosity engine) this mechanism yields a beam pattern following the
prescription of eq.~\ref{eq:stru} under wide conditions. This letter
is organized as follows: in \S~2 we describe the star, cocoon and jet
conditions at breakout; in \S~3 we compute the jet evolution during
the release of the cocoon material and the resulting jet structure at
large radii. We discuss our results and their implications in \S~4 and
we summarize our findings in \S~5.

\newpage

\section{The jet and its cocoon at breakout}

Consider a high entropy mildly relativistic jet ($\Gamma_0 \gtrsim 1$)
generated at some small radius $r_0\approx10^7-10^8$~cm in the core of
a massive star with an opening angle $\theta_0$. The jet tries to
accelerate under the pressure of internal forces. We assume that the
initial jet has enough momentum to define a direction of propagation
and avoid a spherical explosion of the star. Such initial conditions
are analogous to those adopted in numerical simulations (Zhang et
al. 2003, 2004b). In the absence of any external material, it would
accelerate in a broad conical outflow, with its Lorentz factor scaling
with radius as $\Gamma\propto{}r/r_0$.

If the jet is surrounded by dense cold matter, as in the
collapsar/hypernova scenario (Woosley 1993; MacFadyen \& Woosley
1999), such acceleration is inhibited. Once the jet reaches supersonic
speeds with respect to the stellar material, the ram pressure of the
shocked gas ahead of the jet drives a reverse shock into the head of
the jet, slowing the jet head down to subrelativistic speeds. The
mixture of shocked jet and shocked stellar material surrounds the
advancing jet to form a high pressure cocoon, analogous to the cocoon
formed by a low-density jet from a radio galaxy advancing into the
intergalactic medium (Scheuer 1974).  As a result the jet dynamics is
governed by three competing effects. First, the internal pressure
accelerates the jet material, whose Lorentz factor scales as
$\Gamma\propto(\Sigma_j/\Sigma_0)^{1/2}$ if the acceleration is
isentropic, where $\Sigma_j$ is the jet cross section of a single jet
and $\Sigma_0$ its value at $r_0$ (see, e.g., Beloborodov 2003,
\S~3.1). If the jet suffers internal dissipation as it propagates,
e.g., from entrainment of cocoon material or internal shocks, then the
increase of $\Gamma$ with $\Sigma_j$ is slower. This can be caused,
e.g., by shear instabilities at the jet-cocoon boundary.  Second, the
head of the jet is slowed down by its interaction with the massive
star. The larger the jet cross-section, the slower its head
propagates. Both shocked jet and stellar material flow to the sides
feeding the cocoon. Third, the cocoon pressure acts as a collimating
force on the jet, which therefore becomes narrower and more
penetrating, albeit less relativistic.

Matzner (2003) developed a simple analytic treatment to follow this
propagation.  Using the Kompaneets approximation (Kompaneets 1960) to
describe the cocoon expansion, he was able to compute the jet
propagation time within the star and the cocoon properties at the
moment the jet head reaches the stellar surface --- the breakout
time, $t_{\rm br}$. For our purposes we need merely to compare the
width of the cocoon at breakout with the width of the jet.

Consider a jet with luminosity $L_j$ that reaches the stellar surface
at $t_{\rm{br}}$. The energy stored in the cocoon (neglecting
adiabatic losses) is $E_c = L_j\,(t_{\rm{br}}-{r_\star}/{c})$, where
$r_\star$ is the radius of the star.  About half of the cocoon energy
is transferred to the nonrelativistic shocked stellar plasma via the
cocoon shock, with the other half remaining in relativistic, shocked
jet material. If we adopt a relativistic equation of state for the
entire cocoon, we can write the cocoon pressure as
\begin{equation}
p_c \approx \frac{E_c}{3\,V_c}
\label{eq:pc}
\end{equation}
where $V_c\approx{}r_\star\,r_\perp^2$ is the volume and
$r_\perp=v_{\rm{sh}}\,t_{\rm{br}}$ the transverse radius of the cocoon
(we assume that the jet occupies a negligible fraction of the cocoon
volume.) The velocity of the shock $v_{\rm{sh}}$ driven by the cocoon
pressure into the star can be computed by balancing the cocoon
pressure against the ram pressure exerted by the stellar material
(which has negligible internal pressure):
$v_{\rm{sh}}=\sqrt{p_c/\rho_\star}$, where $\rho_\star$ is the matter
density of the star. This set of relations can be used with
eq.~\ref{eq:pc} to obtain an equation for the cocoon pressure that
reads (see also Begelman and Cioffi 1986 for a similar treatment in
the context of extragalactic radio sources):
\begin{equation}
p_c=\left[\frac{L_j\rho_\star}
{3\,r_\star\,t_{\rm{br}}^2}\left(t_{\rm{br}}-\frac{r_\star}{c}\right)
\right]^{1/2}
\simeq\left(\frac{L_j\,\rho_\star}{3\,r_\star\,t_{\rm{br}}}\right)^{1/2} .
\label{eq:pc2}
\end{equation}
We assume that the head of the jet propagates subrelativistically
through the star and write $t_{\rm br} \equiv \eta r_\star/c$, where
$\eta >1$.  Then, adopting the notation $Q=10^x\,Q_x$ and using cgs
units throughout, we have
\begin{equation}
p_c\simeq 3\times10^{19} \eta^{-1/2} \left(
\frac{L_{j,51}\,\rho_{\star}}{r_{\star,11}^2} \right)^{1/2} .
\end{equation}
The opening angle of the cocoon is given by 
\begin{equation}
\theta_c\simeq\frac{r_\perp}{r_\star}\simeq
\frac{\sqrt{p_c/\rho_\star}\,t_{\rm{br}}}{r_\star}
\simeq 0.2 \eta^{3/4} 
\left(\frac{L_{j,51}}{r_{\star,11}^2\,\rho_\star}\right)^{1/4} ,
\label{eq:thc}
\end{equation}
i.e., of the order of tens of degrees.

Now let us consider the properties of the jet as it reaches the
breakout radius. The jet pressure is given by $p_j = L_j/(4 c \Sigma_j
\Gamma_j^2)$, where $\Gamma_j$ is the Lorentz factor of the jet
material before it is shocked at the jet head. Setting $p_j = p_c$
(Begelman \& Cioffi 1989; Kaiser \& Alexnder 1997) with $\Sigma_j =
\pi r_\star^2 \theta_j^2$, we find
\begin{equation}
\theta_{j,{\rm br}} \simeq 0.1 \eta^{1/4} \left(\frac{L_{j,51}}
{r_{\star,11}^2 \rho_{\star}}\right)^{1/4} \Gamma_{j,{\rm br}}^{-1} .
\label{eq:thj}
\end{equation}
For isentropic flow between $r_0$ and $r_\star$ the confined jet
reaches a Lorentz factor
\begin{equation}
\Gamma_{j,{\rm br}}\simeq 14 \eta^{1/8}
\Gamma_0\left(\frac{r_{\star,11}}{r_{0,8}} \right)^{1/2}
\left(\frac{\theta_0}{30^\circ} \right)^{-1/2}
\left(\frac{L_{j,51}}{r_{\star,11}^2 \rho_{\star}}\right)^{1/8} ,
\end{equation}
with somewhat lower values if the jet is dissipative. The
corresponding jet opening angle at breakout is $\la 1^\circ$; however,
the fact that $\theta_{j,{\rm br}}\Gamma_{j,{\rm br}} < 1$ suggests
that the jet should freely expand to a few times $\theta_{j,{\rm br}}$
after exiting the star. Even so, the jet at breakout is expected to be
more than an order of magnitude narrower than the cocoon. A more
detailed treatment of the evolution of the cocoon and its interaction
with the jet can be obtained by numerical integration of the set of
differential equations that govern the flow. This allows us to take
into account the density gradient of the star and the shape of the
cocoon. Results (Lazzati et al. 2005) do not differ substantially from
these analytic estimates.

These results may be altered if substantial dissipation due to
recollimation shocks and/or shear instabilities takes place. We do not
attempt to model these effects here. Recollimation shocks are however
relatively weak, being mostly oblique to the flow. As a matter of
fact, these results are in qualitative agreement with results from
numerical simulations (Zhang, Woosley \& MacFadyen 2003). Simulations
show, indeed, that isentropic conditions are not respected throughout
the propagation, since collimation shocks take place. These shocks
effectively create a new nozzle at a larger radius $r>r_0$ which
modifies the scaling of the Lorentz factor with the jet cross
section. What is of relevance here is that no matter how wide the jet
is initially (and it is likely to be poorly collimated, especially if
initial collimation is provided by the accretion disk), the jet that
emerges from the star is narrow and highly collimated by the cocoon
pressure.  This result is observed in all numerical simulations of
jets in collapsars (MacFadyen \& Woosley 1999; MacFadyen, Woosley \&
Heger 2001; Zhang et al. 2003). The cocoon, on the other hand, covers
a wide solid angle, several tens of degrees across.  In the next
section, we explore what happens after jet breakout.

\section{Jet breakout}

\begin{figure}
\psfig{file=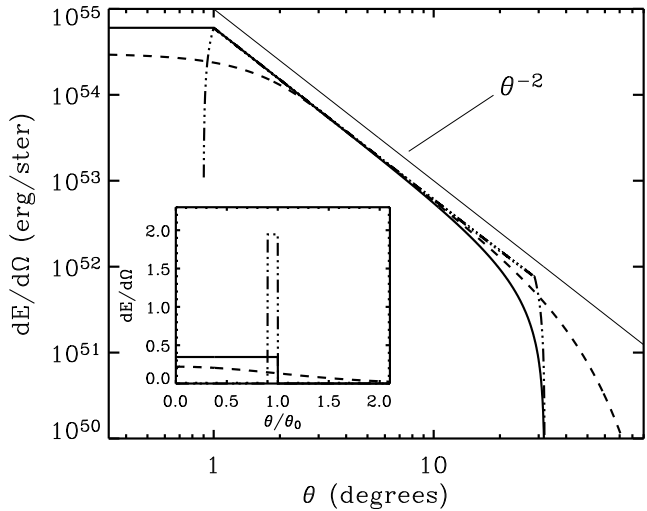,width=\columnwidth}
\caption{{Energy distribution for the afterglow
phase for three instantaneous beam patterns (see inset). In all three
cases a well defined $dE/d\Omega\propto\theta^{-2}$ section is clearly
visible. Only the edge of the jet and its core show marginal
differences. The results shown are for a jet/star with
$L=10^{51}$~erg~s$^{-1}$, $T_{\rm{GRB}}=40$~s and
$r_\star=10^{11}$~cm. Inset: Instantaneous beam patterns that reach
the surface of the star. The solid line shows a uniform jet, dashed
line shows a Gaussian energy distribution, while the dash-dotted line
shows an edge brightened (or hollow) jet.}
\label{fig:pat}}
\end{figure}

We now consider the jet development after the breakout, with the inner
engine still active. At this stage, which has not been investigated in
numerical simulations so far, it is likely that dissipation will have
a lesser role, since the jet, as we shall see, is de-collimated rather
than recollimated.

As the jet reaches the stellar surface, it clears a channel for the
cocoon. The cocoon material is therefore now free to expand out of the
star and its pressure drops. We assume that from this moment on the
shock between the cocoon and the cold stellar material stalls, as a
consequence of the dropping cocoon pressure. This is equivalent to
assuming a constant volume of the cocoon cavity inside the star. The
pressure drop for the relativistic cocoon can be derived through
$dE_c=-\epsilon_c\,\Sigma_c\,c_s\,dt$ where $\Sigma_c$ is the area of
the free surface through which the cocoon material expands,
$\epsilon_c$ the cocoon energy density and $c_s=c/\sqrt{3}$ is the
sound speed of the relativistic gas.  Writing the cocoon volume as
$V_c\sim\Sigma_c\,r_\star$, we can obtain the pressure evolution:
\begin{equation}
p_c=p_{c,{\rm br}}\,\exp\left({-\frac{ct}{\sqrt{3}\,r_\star}}\right)
\end{equation}
where $p_{c,{\rm br}}$ is the cocoon pressure at the moment of shock
breakout.

As the cocoon pressure decreases, fresh jet material passing through
the cocoon is less tightly collimated. Under isentropic conditions the
jet Lorentz factor increases linearly with the opening angle and
pressure balance yields $\theta_j \propto p_c^{-1/4}$, implying an
exponentially increasing opening angle of the form\footnote{Note that,
if the jet is causally connected at breakout as suggested by
eq.~\ref{eq:thj}, the jet would freely expand to an angle
$\theta_j=1/\Gamma_{j,{\rm{br}}}>\theta_{j,{\rm{br}}}$ of the order of
a few degrees. This would result in an initially constant opening
angle. The only effect on the final energy distribution of
eq.~\ref{eq:wow} is to increase the size of the jet core from
$\theta_{j,{\rm{br}}}$ to $1/\Gamma_{j,{\rm{br}}}$.}
\begin{equation}
\theta_j=\theta_{j, {\rm br}}\,\exp\left[\frac{c\,t}{4\sqrt{3}\,r_\star}
\right] .
\label{eq:thopen}
\end{equation}
Dissipative jet propagation gives similar results (with merely a
different numerical coefficient $\sim O(1)$ inside the exponential),
provided that $\Gamma$ varies roughly as a power of $\Sigma_j$.  The
angular distribution of integrated energy, as observed in the
afterglow phase, is computed by integrating the instantaneous
luminosity per unit solid angle from the moment the jet becomes
visible along a given line of sight ($t_{\rm{l.o.s.}}$) until the end
of the burst:
\begin{equation}
\frac{dE}{d\Omega}=\int_{t_{\rm{l.o.s.}}}^{T_{\rm{GRB}}}
\frac{dL}{d\Omega}\,dt \simeq\int_{t_{\rm{l.o.s.}}}^{T_{\rm{GRB}}}
\frac{L(t)}{\pi\,\theta_j^2(t)}\,dt .
\label{eq:struj0}
\end{equation}
$t_{\rm{l.o.s.}}$ is obtained by inverting eq.~\ref{eq:thopen}. Such
integration is valid provided that the jet opening angle at time
$T_{\rm{GRB}}$ is smaller than the natural opening angle of the jet:
$T_{\rm{GRB}}<4\sqrt{3}(r_\star/c)\log(\theta_0/\theta_{\rm{br}})$. For
the fiducial numbers assumed ($r_\star=10^{11}$~cm,
$\theta_{j,\rm{br}}=1\degr$ and $\theta_0=30\degr$) this corresponds to
$\sim100$~s comoving burst duration. Interesting effects should be
expected for longer bursts, as discussed in \S~4. Assuming a jet with
constant luminosity $L$ and for all the lines of sight that satisfy
$t_{\rm br} < t_{\rm{l.o.s.}} \ll T_{\rm GRB}$, eq.~\ref{eq:struj0}
gives the jet structure
\begin{equation}
\frac{dE}{d\Omega} = \frac{2\sqrt{3}\,L\,r_\star}
{\pi\,c}\,\theta^{-2}  \qquad\theta_{j,{\rm{br}}}\le\theta\le\theta_0
\label{eq:wow}
\end{equation}
and $dE/d\Omega\sim$constant inside the core radius
$\theta_{j,{\rm{br}}}$.  This angular dependence, which characterizes
a ``structured jet'' or ``universal jet'' (Rossi et al. 2002, 2004;
Zhang \& Meszaros 2002; Salmonson 2003; Lamb, Donaghy \& Graziani
2003), is of high theoretical interest.  Jets with this beam pattern
reproduce afterglow observations. If the jet is powered by fall-back
of material from the star to the accretion disk, the mass accretion
rate would be anti-correlated with the radius of the star, for a given
stellar mass. This would produce a roughly constant value of
$L_j{}r_\star$ that would reproduce the so-called Frail relation
(Frail et al. 2001; Panaitescu \& Kumar 2001) as a purely
viewing-angle phenomenon.  In this case all jets would be alike, but
different observers would see them from different angles, deriving
different energetics.

In the above equations we have assumed for simplicity that the jet
reaching the surface of the star is uniform. As shown by simulations
(e.g. Zhang et al. 2003), it is more likely that a Gaussian jet
emerges from the star. On the contrary, boundary layers may be
produced by the interaction of the jet with the collimating star,
resulting in edge brightened jets (see the inset of
Fig.~\ref{fig:pat}).  It can be shown easily that the $\theta^{-2}$
pattern does not depend on the assumption of
uniformity. Fig.~\ref{fig:pat} shows the integrated energy
distribution for uniform, Gaussian and hollow intrinsic jets. Even
though small differences are present at the edges (the jet core and
the outskirts), the general behavior is always
$dE/d\Omega\propto\theta^{-2}$.

\section{Discussion}

Our prediction that the jet opening angle evolves with time has two
major, in principle testable, consequences. First, the brightness of a
typical GRB should tend to decrease with time, assuming a constant
efficiency of gamma-ray production, since the photon flux at earth is
proportional to the energy per unit solid angle. This behavior is
certainly seen in FRED (Fast Rise Exponential Decay) single pulsed
events, while it is harder to make a quantitative comparison in the
case of complex light curves. Some events, like the bright GRB~990123,
have multi-peaked lightcurves which show a clear fading at late
times. On the other hand there exist rare events, such as GRB~980923,
in which an increase of luminosity with time is observed. Lacking
afterglow observations, it is hard to determine whether such events
are peculiar.

To understand the second consequence of the model, consider an
observer at an angle $\theta_{\rm{obs}}>\theta_{\rm{br}}$ where
$\theta_{\rm{br}}$ is the jet opening angle at breakout. Initially,
this observer lies outside the beaming cone and does not detect the
GRB. Later, as the jet spreads beyond $\theta_{\rm{obs}}$, the
observer detects the burst. The beginning of the GRB emission is
therefore observer dependent. This causes a correlation between the
burst duration and energetics: the longer the burst the larger its
detected energy output. A correlation between burst duration and
fluence is detected in the BATSE sample but whether it is intrinsic or
due to systematic effects is still a matter of debate. Finally, if any
isotropic emission should mark the jet breakout (MacFadyen \& Woosley
1999; Ramirez-Ruiz, MacFadyen \& Lazzati 2002), such emission would not
be followed immediately by $\gamma$-rays for observers lying outside
of $\theta_{\rm{br}}$. This delay may explain the unusually long
delays between precursors and main emission detected in several BATSE
GRBs (Lazzati 2005).

Some complications can arise for very compact stars and/or very
long-lived jets. In computing the energy profile (eq.~\ref{eq:wow}) we
have implicitly assumed that the jet expands exponentially until it
dies. This approximation obviously fails if the engine is long-lived,
since at some point the jet reaches its natural opening angle --
determined, e.g., by the accretion disk geometry -- and the expansion
stops. Alternatively for some combination of the parameters, the
cocoon may occupy a small opening angle (eq.~\ref{eq:thc}) . As the
jet expands, it eventually hits the cold star material and a different
expansion law, driven by the jet pressure onto the stellar material,
sets in. Again, the expansion of the jet is slowed down and as a
consequence a shallower distribution of energy with off-axis angle is
expected. Such modifications to the $\theta^{-2}$ law would manifest
themselves in the late stages of the afterglow evolution. The higher
than expected energy at large angles would flatten the light curve
decay or produce a bump in the afterglow. Such behavior has been
claimed in several radio afterglows (Frail et al. 2004; Panaitescu \&
Kumar 2004), and could also explain the two breaks observed in the
light curve of GRB~030329\footnote{Note that a simple two-component
jet cannot explain the fast variability observed in the optical
lightcurve. It can only reproduce the overall behavior of the optical
and radio lightcurves.}. If the outflow reaches its natural opening
angle or crashes into the cold stellar material while still active,
the time-integrated jet can be described as the sum of a $\theta^{-2}$
jet plus a roughly uniform jet with a larger opening angle --- in
other words, a two-component jet (Berger et al. 2003). This behavior
is tantalizing also if we consider the interaction from the point of
view of the star. The expansion of the cocoon drives a shock wave into
the cold stellar material where it deposits approximately
$10^{51}$~erg (Zhang et al. 2003; Lazzati \& Begelman in
preparation). If the cocoon is small and the jet comes into contact
with the star while still active and expanding, additional energy is
given to the star. Complex afterglows should therefore be associated
with particularly energetic hypernov\ae, such as SN2003dh.

\section{Summary and conclusions}

In this Letter we have analyzed the evolution of a GRB jet during its
transition from stellar confinement to free expansion. We focused
mainly on the resulting angular structure of the outflow, and reached
two main conclusions. First, the angular structure of the jet in the
afterglow phase does not necessarily reflect the way in which the
energy is released by the inner engine. If the jet opening angle is
not constant during the $\sim100$~s of the engine activity, the
angular structure of the jet results from the integration of this
evolution and has little to do with the pristine beam pattern.
Second, we analyzed the most probable evolution in the collapsar
scenario. We find that the interaction of the jet with the surrounding
cocoon causes the opening angle to increase exponentially with
time. This evolution, for a constant energy release from the inner
engine and independently of the intrinsic beam pattern, produces an
angular structure with $dE/d\Omega\propto\theta^{-2}$.  Such an energy
distribution produces a broken power-law afterglow (Rossi et al. 2002,
2004; Salmonson 2003, Kumar \& Granot 2003; Granot \& Kumar 2003). It
can also explain the so-called Frail relation (Frail et al. 2001;
Panaitescu \& Kumar 2001) if $L_j r_\star/c$ is roughly constant for
different GRBs. The origin of this energy distribution is here
explained for the first time in the context of hydrodynamical
fireballs (see Lyutikov et al. 2003 for a discussion on the context of
electromagnetically dominated fireballs).  We briefly discuss in \S~4
some testable prediction of this model.

\bigskip

We thank Miguel Aloy, Gabriele Ghisellini and Andrew MacFadyen for
useful discussions. This work was supported in part by NSF grant
AST-0307502 and NASA Astrophysical Theory Grant NAG5-12035.

\clearpage

\end{document}